\documentstyle[prd,aps]{revtex}
\def\bra{\langle}
\def\ket{\rangle}

\newcommand{\ba}{\begin{array}}
\newcommand{\ea}{\end{array}}

\begin{document}
\title{The Earth Regeneration Effect of Solar Neutrinos: 
a Numerical Treatment with Three Active Neutrino Flavors}
\author{Jai Sam Kim \footnote[1]{e-mail: jsk@postech.ac.kr}
and Kyungsoo Lee \\
{\it 
Dept of Physics, Pohang University of Science and Technology,\\
Pohang 790-784, Republic of Korea}}
\date{\today}
\maketitle

\begin{abstract}
We introduce an integrated algorithm for full scale investigation of the
earth regeneration effect of solar neutrinos with all three active neutrinos.
We illustrate that the earth effect on low energy solar neutrinos
($E_\nu < 20$ MeV) causes large variations in the survival probabilities 
for large values of the mixing angle $\theta_{12}$ for a given value of
$\theta_{13}$. But it is maximal for small values of $\theta_{13}$ and 
diminishes for large values at a given value of $\theta_{12}$. The nadir 
angle dependence is sensitive to the value of $E'\equiv E/\Delta_{12}$.
As far as low energy solar neutrinos are concerned we notice that the earth 
effect is more pronounced for core penetration. We confirm that the earth 
effect leaves the SMA region almost intact regardless of $\theta_{13}$ and 
that it is significant for high energy $^8$B and $hep$ neutrinos in the LMA 
region. We also show that for large values of $\theta_{13}$, the earth effect 
is significant for high energy neutrinos ($E'\gtrsim 10^9$) and it can be 
large at any nadir angle.

\end{abstract}

\section
{Introduction}
Neutrinos are produced abundantly from thermonuclear reactions taking place
in the stellar core. They carry away most of the energy during the supernova 
explosion. Due to the extreme weakness of its interaction it escapes its
cradle almost intact and provides us with most reliable clues to the internal
physical environments of a star. Thus neutrino astronomy has been established 
as an important research field recently.
The Standard Solar model (SSM) \cite{BP95,BBP98,TUR} predicts the flux of 
the electron neutrino. The historic Homestake experiment \cite{DAV} for
detecting solar neutrinos revealed that the detected neutrinos fall far
short of the SSM prediction.

Several mechanisms were proposed for the depletion of solar neutrinos.
Pontecorvo \cite{PON} proposed the idea of neutrino flavor oscillation earlier 
than any unification theories \cite{GEL}. (see ref. \cite{BIL} for an 
interesting review). The possibility of solar neutrino flux reduction by 
vacuum oscillation was investigated earlier \cite{GRI,FRA}. Subsequent
experiments \cite{HOM,GAL,SAG,KAM} revealed that the solar neutrino deficit 
depends on the energy of the neutrino. Recently the idea of flavor oscillation 
combined with the Mikheyev-Smirnov-Wolfenstein (MSW) effect \cite{WOL,MIK} 
has been most popular. The MSW equations have been vigorously studied.
For simple matter densities, some authors \cite{PAR,TOS,PET,HAX,BEA} derived 
accurate analytic solutions for the two-neutrino MSW evolution equations. 
These solutions have been used for computing the neutrino survival 
probabilities with the approximation of the solar matter density by a mosaic 
of exponential functions. Direct numerical analysis was carried out also 
\cite{FIOR}. Two-flavor MSW solutions to the solar neutrino puzzle were fully 
analysed with high accuracy, including both the day/night effect and spectral 
information \cite{HATA,BKS}. 

Recent Super-Kamiokande experiment \cite{SK} shows an evidence for oscillations
of the atmospheric neutrinos \cite{TH1,TH2,TH3,TH4}. The data are in good 
agreement with two-flavor $\nu_{\mu}\leftrightarrow\nu_{\tau}$ oscillations. 
These results did establish that neutrinos oscillate and possess non-zero 
masses. Thus full three-flavor neutrino oscillation analysis is needed for more 
accurate estimation of neutrino masses and mixing anlges. Even four neutrino
solutions with the sterile neutrino have been suggested \cite{PEL,MAR,JUN,GUN}.

It has been argued \cite{KUO,MIKS,SHI} that under certain conditions the 
solution of the solar neutrino evolution equation in the two neutrino case 
can be extended to the three neutrino case with a slight modification.
There have been many analytic works on three-flavor MSW solutions of the solar 
neutrino problem with simple linear and exponential matter densities
\cite{KUO,MIKS,SHI,GIU,CWK,PTO,ZAG,JOS}.
Most three-flavor MSW fits to the solar neutrino experiments have been made 
within the above framework \cite{HAR,FOG,FOG2,SAK,BAR}.
Recently exact solutions to the three neutrino MSW equations for such simple 
matter densities were derived \cite{TOR,OHL,OSL}. 
Direct numerical analysis is always desirable but it has been hampered by the 
huge computational costs. Recently we presented an efficient numerical 
algorithm for direct computation of the solar neutrino survival probability 
with all three active neutrinos \cite{KIM,JSK}.

According to the PREM model \cite{EAR}, the earth consists of 10 layers of 
different matter densities and compositions. The PREM model gives the matter 
density in each earth layer as a polynomial in radius. It is around 
$\rho=1.6\sim 6.1$ mol/cm$^3$. In such matter densities, the MSW effect can 
be significant for neutrinos with certain energies. The earth regeneration 
effect for solar neutrinos was numerically analysed by many people within the 
two-flavor framework \cite{BOU,CAR,BAL,DAR,CHE,HIR,AUR,SMI,KEA,LOS,BAL2,PET2}.
Since numerical analysis takes a lot of computing time a few attempts to 
derive analytic formula were made. 
Approximating the matter density in each of the five simplified earth layers 
as an even degree quartic polynonimal in radius, Lisi and Montanino \cite{LIS} 
derived analytic solutions for the two-neutrino MSW equations. Other attempts 
were made for three neutrino MSW equations with a periodic step-function 
density profile \cite{AKH} and with constant matter densities \cite{OHL2}.
However, to the best of our knowledge, there are no direct numerical works 
on the earth regeneration effect with all three neutrinos to the depth as 
done in the two flavor case.

In this paper we present an integrated algorithm for analyzing the earth 
regeneration effect of solar neutrinos with all three active neutrinos 
considered. Integration of neutrino evolution equations can be done with 
any ordinary differential equation solver such as adaptive Runge-Kutta or 
Bulirsch-Stoer methods (see, e.g. \cite{BUL}). The wavefunction of a neutrino 
passing through the earth oscillates rapidly with large amplitudes in certain 
ranges of energy and mixing parameters.
The crux of the problem is how to take time averages out of these wildly
oscillating wavefunctions. Two methods have been used in the two neutrino
case. Baltz and Weneser \cite{BAL} carefully analyzed the oscillatory
behavior of the survival probabilities and derived a formula that uses 
instantaneous values of the two particular wavefunctions. Exploiting the 
symmetry of the transition probabilities, Smirnov \cite{SMI} derived a 
formula that uses only one transition probability, $P(\nu_2\rightarrow \nu_e)$.
Due to economy of the Smirnov's formula it has been widely used.
However, we could not extend the formula to the three neutrino case.
There is no shortcut in this case and we need to integrate the evolution
equation for three different initial conditions, we realized. We introduce 
a method to obtain time averages of survival probabilities of solar neutrinos 
after passing through the earth in the three neutrino case. Our method is 
numerical and thus quite general and can be used with any number of flavors.

This paper is organized as follows:
In section 2 we review the basic formalism very briefly. 
In section 3 we introduce our method for taking averages of neutrino fluxes 
as they pass through the earth core.
In section 4 we illustrate the complexity of the earth effect occurring
at large mixing angles. We illustrate how it depends on neutrino energies,
the mixing parameters, and the nadir angles.
In section 5 we present how to make long term averages such as daily
averages in different seasons and monthly and yearly averages.
In section 6 we present sample contour plots for an interesting set of
parameters.

\section
{Basic formalism}
The state of a neutrino can be expressed either in the weak eigenstate basis
$|\nu_\alpha\ket$ or in the mass eigenstate basis $|\nu_i\ket$. The unitary 
matrix $U$ transforms the mass eigenstates into the weak eigenstates,
\begin{equation}
\nu_\alpha = \sum_{i=1}^3 U_{\alpha i} \nu_i,\quad \alpha=e,\mu,\tau .
\end{equation}
The Particle Data Group \cite{PDG} adopts the convention for the mixing matrix,
\begin{equation}
U = \left(
           \begin{array}{ccc}
C_1 C_3 & S_1 C_3 & S_3 \\
-S_1 C_2-C_1 S_3 S_2 & C_1 C_2 - S_1 S_3 S_2 & C_3 S_2 \\
S_1 S_2 - C_1 S_3 C_2 & -C_1 S_2 -S_1 S_3 C_2 & C_3 C_2
           \end{array}
        \right),
\end{equation}
where $C_i\equiv\cos\theta_i$ and $S_i\equiv\sin\theta_i$ and the three 
mixing angles $\theta_1=\theta_{12}$, $\theta_2=\theta_{23}$, and 
$\theta_3=\theta_{13}$ roughly measure mixing between mass eigenstates 
(1-2), (2-3), and (1-3) respectively. We have neglected the CP violating phase, 
which is irrelevant in this problem.

It is convenient to express the state of a neutrino as a mixture of three 
active neutrinos in the weak eigenstate basis,
\begin{equation}
\Phi(t) = a_e |\nu_e\ket + a_\mu(t) |\nu_\mu\ket + a_\tau(t)|\nu_\tau\ket.
\end{equation}
The mass matrix is non-diagonal in the weak eigenstate basis. It is obtained
from the diagonal mass matrix,
\begin{equation}
M_{ij} = \sum_k U_{ik}^\dagger m_k^2 U_{kj}.
\end{equation}
The wavefunctions $a_\alpha(t)$ of a neutrino in a medium obey the evolution 
equations,
\begin{equation}
i\frac{d}{dt} \left(\ba{l} a_e(t) \\ a_\mu(t) \\ a_\tau(t) \ea\right) =
\frac{1}{2\beta}
        \left(\ba{lll}  M_{11} + A_c(t) & M_{12}&  M_{13} \\
                        M_{21} & M_{22}&  M_{23} \\
                        M_{31} & M_{32} & M_{33} \ea\right)
        \left(\ba{l} a_e(t) \\ a_\mu(t) \\ a_\tau(t) \ea\right),
\end{equation}
where $A_c = 2\sqrt{2}G_F E N_e$ is added due to the interaction with the 
charged currents via the MSW mechanism \cite{WOL,MIK} and $\beta=E/R_E$ 
(MeV/cm) and $t=R/R_E$. 

In principle we can use the neutrino evolution equations in either of the two
basis. In practice the mass eigenstate based equations require diagonalization
of the mass matrix at each time step. Diagonalization introduces non-trivial
errors and takes extra computing time in addition to the time to solve the 
ordinary differential equations. Thus we chose the weak eigenstate based 
evolution equations.

Let us make a quick check of the two neutrino resonant condition in the earth 
core where $N_e\simeq 0.466\rho$,
\begin{equation}
\cos(2\theta_{ij})=\sqrt{2}G_F N_e/(\Delta_{ij}/2E),
\end{equation}
while holding the other mixing angles equal to zero. One finds that for a 
small value of the (1-2) mixing angle ($\cos 2\theta_{12}\approx 1$),
the (1-2) transition takes place for neutrinos with 
$E/\Delta_{12}\simeq 10^6\sim 10^7$ eV$^{-1}$. 
For large values of $\theta_{12}$, the resonant condition is satisfied for 
lower energy values. Thus the low energy $pp$-neutrinos can be susceptible 
to the earth matter.

\section
{Computational methods}

The electron neutrinos born from various sources within the solar core come 
out to the solar surface with average probabilities 
[$X_e^2\equiv   P(\nu_e\rightarrow \nu_e)$, 
$X_\mu^2\equiv  P(\nu_e\rightarrow \nu_\mu)$, 
$X_\tau^2\equiv P(\nu_e\rightarrow \nu_\tau)$].
In the interesting ranges of mixing parameters, $\Delta_{12}$, $\theta_{12}$, 
$\theta_{13}$, we have compiled a large set of solar neutrino survival 
probabilities born at selected radial positions $r$ with selected energies 
$E$ \cite{KIM,JSK}. We used $\Delta_{23}=2.2\times 10^{-3}$ eV$^2$ and 
$\theta_{23}=43.5^\circ$.

Integration of the neutrino evolution equations in the earth are carried out 
using the CERN library \cite{CERN} routine DDEQBS, which is an implementation 
of the Bulirsch-Stoer algorithm. We used the PREM model \cite{EAR} with 
$N_e=0.466\rho$ in the core ($R<5463$km) and $N_e=0.494\rho$ in the mantle. 
Our results agree with previous works \cite{CAR,BAL,OHL2}. We especially 
checked our results against those of \cite{OHL2} and found good agreements. 
(see Fig. 1a). 
It would be extremely time-consuming to trace all these neutrinos individually 
as they pass through the earth. Instead we trace how three pure neutrinos, 
$\Psi_e(t=0)=(1,0,0,0,0,0)$, $\Psi_\mu(t=0)=(0,0,1,0,0,0)$ and 
$\Psi_\tau(t=0)=(0,0,0,0,1,0)$ evolve as they pass through the earth for a 
given set of parameters, $\Delta_{12}$, $\Delta_{23}$, $\theta_{12}$, 
$\theta_{13}$, $\theta_{23}$. We compile the wavefunctions for selected values 
of the energy $E$ and the zenith angle $\eta$. Then we set up a handy 
interpolation function to be used for computing the neutrino fluxes arriving 
at various experiments in any given period of time.

Let us denote the states of pure neutrinos after passing through the earth as,
\begin{eqnarray}
\Psi_e(t) & = & 
 \alpha_e(t)|\nu_e\ket
+\alpha_\mu(t)|\nu_\mu\ket
+\alpha_\tau(t)|\nu_\tau\ket, \nonumber\\
\Psi_\mu(t) & = & 
 \beta_e(t)|\nu_e\ket
+\beta_\mu(t)|\nu_\mu\ket
+\beta_\tau(t)|\nu_\tau\ket, \\
\Psi_\tau(t) & = & 
 \gamma_e(t)|\nu_e\ket
+\gamma_\mu(t)|\nu_\mu\ket
+\gamma_\tau(t)|\nu_\tau\ket. \nonumber
\end{eqnarray}
Then the state of a solar neutrino in a mixed state with probabilities, 
$(X_e^2,X_\mu^2,X_\tau^2)$ can be written as a linear sum of three pure states, 
\begin{equation}
\Psi_{SE}(t) 
 = X_e\exp(i\phi_1) \Psi_e(t) + X_\mu\exp(i\phi_2) \Psi_\mu(t) 
+ X_\tau \Psi_\tau(t),
\end{equation}
where we assumed some phases for solar neutrinos while keeping the phase of 
the $\tau$ neutrino zero.

Let us denote the state of a solar neutrino arriving at the earth as,
\begin{equation}
\Psi_S(t=0) \equiv a_e|\nu_e\ket + a_\mu |\nu_\mu\ket + a_\tau |\nu_\tau\ket.
\end{equation}
Then the probability of a solar neutrino to emerge as an electron neutrino
after passing though the earth is written as
\begin{eqnarray}
P_{SE} 
& = & |\bra\nu_e|\Psi_{SE}\ket|^2 = 
|a_e\alpha_e+a_\mu \beta_e+a_\tau \gamma_e|^2 \nonumber \\
& = & |a_e|^2 |\alpha_e|^2+|a_\mu|^2 |\beta_e|^2+|a_\tau|^2 |\gamma_e|^2 \nonumber \\
& &
+[a_e\alpha_e a_\mu^*\beta_e^* + {\rm c.c.}] 
+[a_\mu\beta_e a_\tau^*\gamma_e^* + {\rm c.c.}] 
+[a_\tau\gamma_e a_e^*\alpha_e^* + {\rm c.c.}]. 
\end{eqnarray}

In order to obtain the time average of $P_{SE}(t)$ it is more convenient to
use the mass eigenstate basis. Let us denote
\begin{eqnarray}
\alpha_e(t) & = & s_1 e^{i\omega_1 t}+s_2 e^{i\omega_2 t}+s_3 e^{i\omega_3 t},\\
 \beta_e(t) & = & t_1 e^{i\omega_1 t}+t_2 e^{i\omega_2 t}+t_3 e^{i\omega_3 t},\\
\gamma_e(t) & = & u_1 e^{i\omega_1 t}+u_2 e^{i\omega_2 t}+u_3 e^{i\omega_3 t},
\end{eqnarray}
where $s_i$, $t_i$, $u_i$ are independent of time. Then we have
\begin{equation}
\alpha_e \alpha_e^* = s_1s_1^*+s_2s_2^*+s_3s_3^*+{\rm cross\ terms},
\end{equation}
\begin{eqnarray}
\alpha_e \beta_e^* &=& s_1t_1^*+s_2t_2^*+s_3t_3^*+{\rm cross\ terms} \nonumber\\
&=& (s_{1R}t_{1R}+s_{1I}t_{1I})+i(s_{1I}t_{1R}-s_{1R}t_{1I})
   +(s_{2R}t_{2R}+s_{2I}t_{2I})+i(s_{2I}t_{2R}-s_{2R}t_{2I}) \nonumber \\
&+&(s_{3R}t_{3R}+s_{3I}t_{3I})+i(s_{3I}t_{3R}-s_{3R}t_{3I})+{\rm cross\ terms}\\
&\equiv&R_{\alpha\beta}+i I_{\alpha\beta} + {\rm cross\ terms},\nonumber
\end{eqnarray}
where 
\begin{eqnarray}
R_{\alpha\beta} &=& (s_{1R}t_{1R}+s_{1I}t_{1I})+(s_{2R}t_{2R}+s_{2I}t_{2I})+
(s_{3R}t_{3R}+s_{3I}t_{3I}), \\
I_{\alpha\beta} &=& (s_{1I}t_{1R}-s_{1R}t_{1I})+(s_{2I}t_{2R}-s_{2R}t_{2I})+
(s_{3I}t_{3R}-s_{3R}t_{3I}).
\end{eqnarray}
The cross terms cancel out upon time averaging.

It is convenient to express the amplitudes $(a_e,a_\mu,a_\tau)$ in the weak
eigenstate basis in terms of those in the mass eigenstate basis,
$(a_1,a_2,a_3)$,
\begin{eqnarray}
a_e(t) &=& U_{11} a_1(t) + U_{12} a_2(t) + U_{13} a_3(t), \nonumber\\
a_\mu(t) &=& U_{21} a_1(t) + U_{22} a_2(t) + U_{23} a_3(t), \\
a_\tau(t) &=& U_{31} a_1(t) + U_{32} a_2(t) + U_{33} a_3(t). \nonumber
\end{eqnarray}
Averaging over time, we obtain
\begin{equation}
X_e^2 = \bra a_e a_e^* \ket =
U_{11}^2 |a_1|^2 + U_{12}^2 |a_2|^2 + U_{13}^2 |a_3|^2, \cdots.
\end{equation}
We solve this equation for $|a_i|^2$. Then we can express 
$\bra a_e a_\mu^* \ket$ in terms of $|a_i|^2$. Let us denote
\begin{equation}
\bra a_e a_\mu^* \ket = 
U_{11} U_{21} a_1 a_1^* + U_{12} U_{22} a_2 a_2^* + U_{13} U_{23} a_3 a_3^* 
\equiv X_{e\mu}+iY_{e\mu}.
\end{equation}
We notice that the phases of $a_i$ become irrelevant upon time-averaging.
Now we get the time average
\begin{equation}
\bra [a_e\alpha_e a_\mu^*\beta_e^* +{\rm c.c.}]\ket = 
[(R_{\alpha\beta}+i I_{\alpha\beta})(X_{e\mu}+iY_{e\mu}) +{\rm c.c.}]
= 2(R_{\alpha\beta}X_{e\mu} - I_{\alpha\beta}Y_{e\mu}).
\end{equation}
Similarly working on other terms and collecting all terms, we obtain the
time averaged probability of a solar neutrino to emerge as an electron neutrino
after passing the earth,
\begin{eqnarray}
\bra P_{SE} \ket &=&
|a_e|^2 |\alpha_e|^2+|a_\mu|^2 |\beta_e|^2+|a_\tau|^2 |\gamma_e|^2  \nonumber\\
&+& 2(R_{\alpha\beta}X_{e\mu} - I_{\alpha\beta}Y_{e\mu}
+ R_{\beta\gamma}X_{\mu\tau} - I_{\beta\gamma}Y_{\mu\tau}
+ R_{\gamma\alpha}X_{\tau e} - I_{\gamma\alpha}Y_{\tau e}).
\end{eqnarray}

In order to check our algorithm we reproduced some figures published in the 
literatures \cite{BAL,LIS} in the limit $\theta_{13}\rightarrow 0$. 
(see Fig. 1). Our figures agree with the results as illustrated in Fig. 4 of 
\cite{BAL} in the range $E<10^8$. 
Our figures completely agree with the results obtained by using the time 
averaged $P_E(\nu_2\rightarrow \nu_e)$ in Eqn. (1) of \cite{LIS}.

\section
{Survival probabilities of a solar neutrino with non-zero 
$\boldmath\Large\theta_{13}$}

We consider the effects of non-zero $\theta_{13}$ on the electron neutrino 
survival probabilities of an electron neutrino created at the center of the 
Sun as a function of $E'\equiv E/\Delta_{12}$ in MeV/eV$^2$. We illustrate 
two cases in Figs. 2. In Fig. 2a we consider the Small Mixing Angle (SMA) case 
using the set of parameter values, ($\sin^2(2\theta_{12})=4.40\times 10^{-3}$,
$\Delta_{12}=6.31\times 10^{-6}$ eV$^2$). It consists of 3 plots corresponding 
to $\theta_{13}=5^\circ,\ 22.5^\circ, \ 45^\circ$. 
Fig. 2b corresponds to the Large Mixing Angle (LMA) case and
($\sin^2(2\theta_{12})=0.76$, $\Delta_{12}=1.8\times 10^{-5}$ eV$^2$) was used.
The upper limit of solar neutrino energy, 20 MeV is indicated by the downward 
arrow. Each plot contains a curve for raw solar neutrino flux and five curves 
marked by Sm corresponding to five neutrino paths with nadir angles 
$\eta=(m/10)(\pi/2)$.

Neutrinos undergo resonance conspicuously in those regions where the curves 
oscillate widely away from the raw curve. We notice that for nonzero values of 
the (1-3) mixing angle $\theta_{13}$ double resonances occur, one in the range 
$10^{6}\lesssim E'\lesssim 10^{7}$ and the other in
$10^{8}\lesssim E'\lesssim 10^{10}$.
The resonance in the low range is caused by non-zero $\theta_{12}$ and its 
amplitude grows larger for large $\theta_{12}$. The resonance in the high 
range is due to non-zero $\theta_{13}$ and it is enhanced for large 
$\theta_{13}$. The resonances occur for all nadir angles with almost equal 
strengths. It is noteworthy that for low energy solar neutrinos the earth 
effect is larger for smaller values of the nadir angle.

Fig. 2a reveals that in the SMA case the earth effect for low energy solar 
neutrinos ($E<20$ MeV) is minor for all values of $\theta_{13}$.
Fig. 2b exhibits that in the LMA case the earth effect is non-trivial for some 
high energy $^8$B and $hep$ solar neutrinos. It is most significant for small 
values of $\theta_{13}$ and diminishes for large $\theta_{13}$.

Since it costs a lot of computation to trace neutrinos for each nadir angle
we need to make a strategy for handling the angular effect. We exploit that
for a given set of parameters, the neutrino wavefunctions are smooth functions
of the nadir angle $\eta$. We thus compute at a finite number of angles and
then build an interpolation function. For the nadir angle, we choose 
$\cos(\eta)$ as the sampling function. Thus we sample more frequently near 
$\eta=0$ and scarcely near $\eta=\pi/2$. We selected 25 nadir angles between 
0 and $\pi/2$. For low energy solar neutrinos the earth effect diminishes 
for $\eta > \eta_c=0.5779$, the nadir angle grazing the boundary between the 
core and the mantle. Thus our sampling strategy works fine. An interpolation 
function made from these 25 $\eta$ values will be fairly accurate to use.

We can now compute the states of pure neutrinos, $\Psi_e(t)$, $\Psi_\mu(t)$, 
$\Psi_\tau(t)$ at discrete energies and nadir angles selected according to 
the importance sampling functions. In our previous works on solar neutrino 
flux \cite{KIM}, we selected energy sampling points according to the 
convolution of the neutrino energy distribution functions as they are created 
in the solar core and the detector cross sections. In the present work we 
added 7 more energy sample points in the interval $0.23\leq E \leq 0.737$ MeV 
to account for the CNO neutrinos more faithfully. 

\section
{Long term averages: Daily, Monthly and Yearly averages}

We have compiled a large amount of raw solar neutrino fluxes for a reasonably 
large area in the parameter space, $(\Delta_{12},\sin^2(2\theta_{12}))$ for 
selected values of $\theta_{13}$, using the method described in \cite{KIM}. 
Our results were partially reported without the earth effect in \cite{JSK}. 
The results of our computations in the $\theta_{13}\rightarrow0$ limit have 
been carefully checked against the results obtained with the analytic 
approximation formula such as the Petcov's improved formula \cite{PET}.
We agree with their results in the range of creation positions where their 
results are reliable as they claimed, in the sense that their formula give 
the flux values averaged over the creation positions whereas ours give the 
fluxes created at exact positions as illustrated in Fig. 5 of Ref. \cite{TOR}. 
We also compared the survival probabilities of a neutrino created at the
center of the Sun in the range of $E'$ for various sets of mixing parameters.
The agreement is to one part in 1000 in the worst case. Since our results are 
computed numerically we have equally accurate fluxes beyond their claimed 
range. For three neutrino cases, analytic formula for survival probabilities 
for simple densities have been derived only recently \cite{TOR,OHL,OSL}. We 
have already confirmed that our subroutines reproduce the same results as 
those given in \cite{OHL2}.

Next we compute seasonal variations of daily averages. The nadir angle $\eta$ 
of a neutrino path can be written \cite{LIS} as a function of the date of year 
$\tau_d$ from the winter solstice, the time of day $\tau_h$ from the midnight 
and the lattitude $\lambda$ as follows,
\begin{equation}
\cos(\eta) = 
\cos(\lambda)\cos(\tau_h)\cos(\delta_S) -\sin(\lambda)\sin(\delta_S)
\end{equation}
where the Sun declination is given by $\sin(\delta_S)=-\sin(i_E)\cos(\tau_d)$ 
and $i_E=0.4091$ is the earth inclination.

We illustrate the survival probabilities for two interesting parameter sets
in Figs. 3. The average was taken over every 12 minuites for 24 hours on
three days, the winter solstice (dotted), the vernal equinox (dash-dotted)
and the summer solstice (dashed). The raw solar neutrino flux is plotted with 
the solid curve. The lattitude of Gran Sasso was used.
The downward arrows mark the point of $E=20$ MeV, the upper bound of the solar
neutrino energy. Each set consists of 4 plots corresponding to
$\theta_{13}=0^\circ,\ 1.5^\circ,\ 15^\circ,\ 22.5^\circ$.
As we notice from Fig. 3a, in the small mixing angle case the electron neutrino 
survival probability is almost unaffected by the earth effect as in the two 
neutrino case, regardless of $\theta_{13}$. Fig. 3b shows that in the large 
mixing angle case (we used $\sin^2(2\theta_{12})=0.63$, 
$\Delta_{12}=1.3\times 10^{-5}$ eV$^2$ in Fig. 3b) the earth effect is 
non-negligible for high energy neutrinos. The effect is maximal for 
$\theta_{13}=0$ and diminishes as $\theta_{13}$ is increased.

We are now equipped with all the routines that are needed to compute average 
fluxes in an extended period of time. Since we used the importance sampling 
method for selecting reasonably large number of creation positions and energies 
to set up an interpolation function that is particularly good in the important 
ranges, we can reproduce the fluxes with high accuracy for creation position 
and energy values within the those ranges.

We have computed a large number of states of pure neutrinos,
$\Psi_e(t)$, $\Psi_\mu(t)$, $\Psi_\tau(t)$ at 100 discrete energies and 25 
nadir angles. We used 40 energy sample points for $pp$ neutrinos in the 
interval $0.23<E<0.737$ MeV similarly picked as in \cite{KIM,JSK} and 
the same 60 points as used in \cite{KIM,JSK} in the interval $0.737<E<19$ MeV 
for boron and CNO neutrinos, for the same set of parameters 
$(\Delta_{12},\sin^2(2\theta_{12}),\theta_{13})$.
We handpicked three important energy points, $E=0.384,\ 0.862,\ 1.442$ MeV
for Be neutrinos and pep neutrinos repectively. Each integration of the 
neutrino evolution equations in the earth takes only a modest amount of 
computing time but the huge number of integrations requires a large amount of 
total computing time. We again have to resort to a parallel supercomputer. The 
parallelization can be done using the same method as described in \cite{KIM}.

Using these data we can compute the earth regeneration effect on the survival 
probability of an incident solar neutrino at any energy and any nadir angle
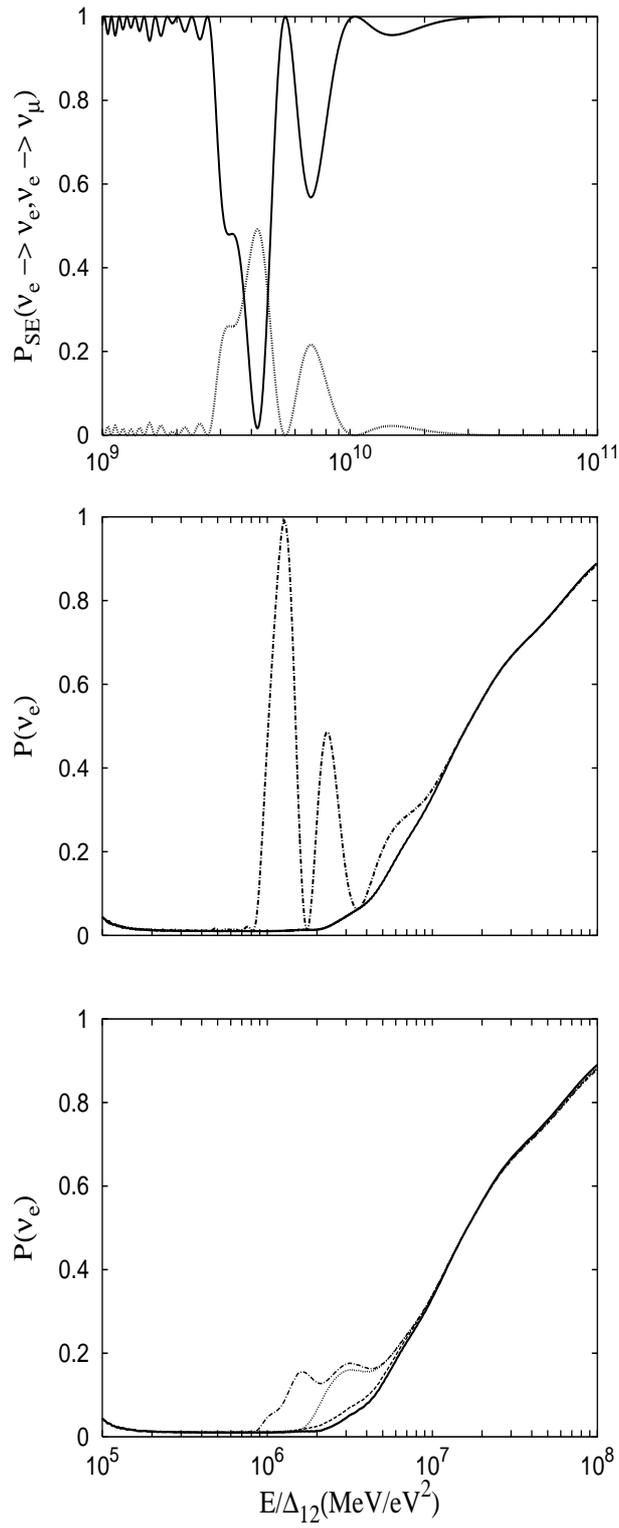
for a selected set of mixing parameters. In this way we can compute the earth 
regenerated solar neutrino fluxes for a large region of the mixing parameter 
space with a modest number of energy samplings to yield an accurate estimate 
of event rates.

Some cautions are needed for setting up an interpolation function for
creation positions and energies. We illustrate the solar neutrino survival
probability as a function of creation position in Fig. 4, where we see a very 
sharply rising curve. We see a sharply dropping curve in Fig. 2a. Since we 
are using only a finite number of samples with different spacings we will 
encounter a sticky situation when the curves happen to turn sharply in the 
region where the sample spacing is wide. The popular cubic spline interpolation 
fails badly in this situation. One needs to use the exponential spline 
\cite{BUL}. We used the FITPACK subroutines available in the GAMS library 
\cite{NET}. As Fig. 4 shows the interpolated curve is as good as the original 
one.

\section
{Sample Contour plots}
We illustrate sample iso-SNU(FLUX) contour plots in the plane,
($\sin^2(2\theta_{12})-\Delta_{12}$),
for a set of mixing parameters,
$\theta_{13}=10^\circ$, $\Delta_{23}=2.2\times 10^{-3}$ eV$^2$, and
$\theta_{23}=43.5^\circ$ in Fig. 5. For a given set of 
($\sin^2(2\theta_{12}),\Delta_{12}$) we had generated solar neutrinos 
according to the SSM model \cite{BP95,BBP98}. In order to compute the raw 
solar neutrino fluxes we used 36 creation positions and 60 energy samples 
for $pp$ neutrino and 60 creation positions and 60+7 energy samples for others.
For the same set of mixing parameters ($\sin^2(2\theta_{12}), \Delta_{12}$) 
we had computed the states of three pure neutrinos,
$\Psi_e(t)$, $\Psi_\mu(t)$, $\Psi_\tau(t)$ at 100 discrete energies and 25
nadir angles optimally selected as explained in section IV. For each solar 
neutrino created at position $r$ with energy $E$ we perform a sequence of time 
averaging operations leading to Eq. (22). In this way we build an interpolation 
function for $\bra P_{SE}(r,E,\eta)\ket$ as a function of creation position 
$r$, energy $E$ and nadir angle $\eta$. We integrate the fluxes over the 
creation positions according to the relevant creation probability distribution 
functions and then build an interpolation function for 
$\bra P_{SE}(E,\eta) \ket$ as a function of the energy and nadir angle. 
We can now take seasonal and yearly averages of this flux to get 
$\bra P_{SE}(E) \ket$, which involves integration over $\eta$ with a definite
lattitude $\lambda$. Finally we take a convolution of $\bra P_{SE}(E) \ket$ 
with the absorption cross sections $\sigma(E)$ to get the yearly average of 
event rate for a particular experiment.

In order to make iso-SNU(FLUX) contour plots, we used a $16\times 21$ 
logarithmic mesh in the ($\sin^2(2\theta_{12})-\Delta_{12}$) plane in the 
range $0.001\leq \sin^2(2\theta_{12})\leq 1$ and 
$1.58\times 10^{-6} \leq \Delta_{12} \leq 1.58\times 10^{-4}$ eV$^2$.
We padded two extra values, (0.85,0.97), for $\sin^2(2\theta_{12})$ in order
to examine the large $\theta_{12}$ region more carefully. 

We made separate plots for each experiment \cite{HOM,GAL,SAG,KAM} with 
appropriate lattitude $\lambda$. As expected from previous figures, contours 
in the SMA region are almost intact but those in the small 
$\Delta_{12}< 10^{-5}$ eV$^2$ and large $\sin^2(2\theta_{12}) > 0.1$ region 
are distorted significantly by the earth effect. But for gallium detectors 
the curves are not distorted visibly in the considered region of parameter
space. This is due to the fact that the major signal source for gallium 
detectors is the $pp$ neutrinos with very low energies, for which the earth 
regeneration effect is minimal.
The overall earth effect for solar neutrinos is larger
for small $\theta_{13}$ than for large $\theta_{13}$ as we see from Fig. 6.

\section
{Summary and Conclusion}
We introduced an integrated algorithm to deal with the earth regeneration
effect of solar neutrinos with all three active neutrinos considered.
Main ingredients of the algorithm are the time averaging algorithm with the 
use of mass eigenstates, the strategy for sampling energy and nadir angle, 
and the interpolation algorithm. Our algorithm is useful for full scale 
investigation of the earth effect.

We illustrated that the earth effect on low energy solar neutrinos causes
large variations in the survival probabilities for large mixing angles
$\theta_{12}$ at a given value of $\theta_{13}$. But it is maximal for small 
values of $\theta_{13}$ and diminishes for large values ($\sim 45^\circ$) 
at a given value of $\theta_{12}$. The nadir angle dependence is sensitive to 
the value of $E'$. As far as low energy solar neutrinos are concerned we 
notice that the earth effect is more pronounced for smaller nadir angles.

We have shown that for large values of $\theta_{13}$, the earth effect is 
significant for high energy accelerator or atmospheric neutrinos 
($E'\gtrsim 10^9$) and it can be large at any nadir angle.

This work was funded by POSTECH research fund.

\end{document}